\documentclass[oribibl]{llncs}
\usepackage{etex}

\usepackage[utf8]{inputenc}
\usepackage[T1]{fontenc}
\usepackage{cite}
\usepackage{amsmath}
\usepackage{amssymb}
\usepackage{mathtools}
\usepackage{hyperref}
\usepackage{caption}
\usepackage{subcaption}
\usepackage{tikz}
\usepackage{pgfplots}
\usepackage{xspace}
\usepackage{booktabs}
\usepackage{multirow}
\usepackage{listings}
\usepackage{float}

\usepackage{aicescover}

\colorlet{graybg}{gray!10}
\colorlet{plot1}{red!75!white}
\colorlet{plot2}{green!75!black}
\colorlet{plot3}{blue}
\colorlet{plot4}{yellow!75!black}
\colorlet{plot5}{violet}
\colorlet{plot6}{cyan}
\colorlet{plot7}{orange}
\colorlet{plot8}{black}

\usetikzlibrary{
    external,
    scopes,
}

\pgfkeys{
    /tikz/external/mode=list and make
}
\tikzsetexternalprefix{figures/externalized/}
\tikzset{external/export=false}

\pgfplotsset{
    every axis/.append style={
        axis background/.style={fill=graybg},
        legend cell align=left,
        xlabel near ticks,
        ylabel near ticks,
        enlarge x limits={value=0.05, auto},
        enlarge y limits={value=0.05, auto},
        width=\textwidth,
        height=.5\textwidth,
        ymajorgrids=true,
        xmin=0,
    },
    twocolplot/.style={
        height=.8\textwidth,
    },
    every axis plot/.append style={
        thick
    },
    perfplot/.style={
        ylabel={flops/cycle},
        ymin=0,
    },
    predplt/.style={
        perfplot,
        cycle list name=predplt,
        xlabel={$a = b = c$ \quad $(i = 8)$},
    },
    plotdot/.style={plot1},
    plotaxpy/.style={plot2},
    plotgemv/.style={plot3},
    plotger/.style={plot4},
    plotgemm/.style={plot5},
}
\pgfplotscreateplotcyclelist{predplt}{
    {plot1,dashdotted},
    {plot2,dashdotted},
    {plot3,dashdotted},
    {plot4,dashdotted},
    {plot5,dashdotted},
    {plot6,dashdotted},
    {plot1,dashed},
    {plot2,dashed},
    {plot3,dashed},
    {plot4,dashed},
    {plot5,solid},
    {plot6,solid},
}
\pgfplotscreateplotcyclelist{default}{
    {plot1},
    {plot2},
    {plot3},
    {plot4},
    {plot5},
    {plot6},
    {plot7},
    {plot8},
}

\lstdefinelanguage{algs}{
    keywords=[1]{for},
    keywordstyle=[1]{\bf},
}
\hypersetup{
    colorlinks=true,       
    linkcolor=blue!50!black,          
    citecolor=green!50!black,        
    filecolor={.5 0 0},      
    urlcolor=red!50!black,           
}

\newfloat{algorithms}{t}{loa}
\floatname{algorithms}{Algs.}

\newcommand\algname[2]{$#1$-{\tt#2}\xspace}
\renewcommand\dot{{\tt dot}\xspace}
\newcommand\axpy{{\tt axpy}\xspace}
\newcommand\gemv{{\tt gemv}\xspace}
\newcommand\ger{{\tt ger}\xspace}
\newcommand\gemm{{\tt gemm}\xspace}

\newcommand\pluseq{\mathrel{{+}{=}}}

\newenvironment{algorithm}[2]{
    \vspace{.5\baselineskip}

    \newcommand\loops{#1}
    \newcommand\kernel{#2}
    \tikzexternaldisable
    \begin{tikzpicture}
        \node[anchor=north, draw, fill=graybg] (alg) \bgroup%
            \begin{minipage}{.6\textwidth}%
}{%
            \end{minipage}
        \egroup;
        \node[anchor=north east, draw, fill=white, inner ysep=1.5pt] at (alg.north east) {
            \scriptsize
            \algname{\loops}{\kernel}\ifcsname contraction\endcsname(\ref*{plt:\contraction:\loops_\kernel})\else\fi
        };
    \end{tikzpicture}
    \tikzexternalenable
}

\newenvironment{drawcube}[1][(0,0,0)]{
    \begin{scope}[
        shift={#1},
        inner/.style={plot1, thick, line join=round, line cap=round, fill=plot1, fill opacity=.25},
        outer/.style={plot3, line join=round, fill=plot3, fill opacity=.1},
        scale=.5
    ]
        \newcommand\slicenone[1][inner]{\draw[##1] 
            +(-1,-1,1) rectangle +(1,1,1) -- +(1,1,-1)
            +(-1,1,1) -- +(-1,1,-1) -- +(1,1,-1) -- +(1,-1,-1) -- (1,-1,1);
        }
        \newcommand\sliceA[1][inner]{\draw[##1] +(-1,0,-1) -- +(1,0,-1) -- +(1,0,1) -- +(-1,0,1) -- cycle;}
        \newcommand\sliceB[1][inner]{\draw[##1] +(0,-1,-1) -- +(0,1,-1) -- +(0,1,1) -- +(0,-1,1) -- cycle;}
        \newcommand\sliceC[1][inner]{\draw[##1] +(-1,-1,0) rectangle +(1,1,0);}
        \newcommand\sliceAB[1][inner]{\draw[##1] +(0,0,-1) -- +(0,0,1);}
        \newcommand\sliceAC[1][inner]{\draw[##1] +(-1,0,0) -- +(1,0,0);}
        \newcommand\sliceBC[1][inner]{\draw[##1] +(0,-1,0) -- +(0,1,0);}
        \newcommand\sliceABC[1][inner]{\draw[##1, ultra thick] +(0,0,0) -- +(0,0,0);}
        \draw
            +(-1,-1,-1) -- +(1,-1,-1)
            +(-1,-1,-1) -- +(-1,1,-1)
            +(-1,-1,-1) -- +(-1,-1,1);
}{
        \slicenone[]
    \end{scope}
}

\newenvironment{drawsquare}[1][(0,0,0)]{
    \begin{scope}[
            shift={#1},
            inner/.style={plot1, thick, line join=round, line cap=round, fill=plot1, fill opacity=.25},
            outer/.style={plot3, line join=round},
            scale=.5
        ]
        \newcommand\slicenone[1][inner]{\draw[##1] +(-1,-1,0) rectangle +(1,1,0);}
        \newcommand\sliceA[1][inner]{\draw[##1] +(-1,0,0) -- +(1,0,0);}
        \newcommand\sliceB[1][inner]{\draw[##1] +(0,-1,0) -- +(0,1,0);}
        \newcommand\sliceAB[1][inner]{\draw[##1, ultra thick] +(0,0,0) -- +(0,0,0);}
        \draw +(-1,-1,0) rectangle +(1,1,0);
}{
    \end{scope}
}

\newcommand\parsum[1]{\textcolor{orange}{#1}}
\renewcommand\parsum[1]{}

\title{
    On the Performance Prediction of\texorpdfstring{\\}{}
    BLAS-based Tensor Contractions
}
\author{
    Elmar Peise, Diego Fabregat-Traver, and Paolo Bientinesi
}
\institute{
    AICES, RWTH Aachen\\
    \email{\{peise,fabregat,pauldj\}@aices.rwth-aachen.de}
}

\aicescovertitle{
    On the Performance Prediction\texorpdfstring{\\}{}
    of BLAS-based\texorpdfstring{\\}{}
    Tensor Contractions
}

\begin{document}
     \aicescoverpage
    \maketitle

    \begin{abstract}
    Tensor operations are surging as the computational building blocks for a
    variety of scientific simulations and the development of high-performance
    kernels for such operations is known to be a challenging task.  While for
    operations on one- and two-dimensional tensors there exist standardized
    interfaces and highly-optimized libraries (BLAS), for higher dimensional
    tensors neither standards nor highly-tuned implementations exist yet.  In
    this paper, we consider contractions between two tensors of arbitrary
    dimensionality and take on the challenge of generating high-performance
    implementations by resorting to sequences of BLAS kernels.  The approach
    consists in breaking the contraction down into operations that only involve
    matrices or vectors.  Since in general there are many alternative ways of
    decomposing a contraction, we are able to methodically derive a large family
    of algorithms.  The main contribution of this paper is a systematic
    methodology to accurately identify the fastest algorithms in the bunch, {\em
    without executing them}.  The goal is instead accomplished with the help of
    a set of cache-aware micro-benchmarks for the underlying BLAS kernels.  The
    predictions we construct from such benchmarks allow us to reliably single
    out the best-performing algorithms in a tiny fraction of the time taken by
    the direct execution of the algorithms.
\end{abstract}

\parsum{orange: paragraph summary}


    \section{Introduction}
    \label{sec:intro}
    \parsum{origin of tensor contractions}
Tensor contractions play an increasingly important role in various scientific
computations such as general relativity and electronic structure calculations in
quantum chemistry.  Computationally, contractions are generalizations of
matrix-vector and matrix-matrix products that involve operands of higher
dimensionality.  While there are several highly-tuned implementations of the
Basic Linear Algebra Subprograms (BLAS)~\cite{blas1, blas2, blas3} for operands
with up to 2 dimensions, there are no equivalently standardized high-performance
libraries for general tensor contractions.  Fortunately, just as matrix-matrix
products can computationally be decomposed into a sequence of matrix-vector
products, most higher dimensional tensor contractions can be cast in terms of
matrix-matrix or matrix-vector BLAS kernels.  However, each tensor contraction
can be computed via BLAS kernels in many, even hundreds, of different ways, each
with its own performance signature.  This work addresses the problem of
accurately predicting the performance of BLAS-based algorithms for tensor
contractions. 

One could argue that only algorithms that use the \gemm kernel\footnote{%
    \gemm is the BLAS-3 routine for matrix-matrix multiplication, which on many
    systems is optimized within a few percent of peak performance.
} are real candidates to achieve the best performance; while for the most part
this observation is true, due to the fact that in practical contractions it is
often the case that one or more dimensions are very small (while BLAS is mostly
optimized for large dimensions), the difference in performance between two
\gemm-based algorithms can be dramatic.  At any rate, with this work we aim at
the accurate prediction of any BLAS-based contraction, irrespective of which
kernel is used.  Our approach, which never resorts to timing a full algorithm,
makes use of what we call micro-benchmarks.  These are benchmarks that only
execute one BLAS operation in a prescribed memory environment. The idea is to
analyze the structure of the code, and determine the status of the cache
(precondition) prior to the execution of the kernel; we recreate carefully such
status within the micro-benchmark so that the specific kernel can be timed in
conditions analogous to those experienced in the actual algorithm.  Based on
these timings, we extrapolate the total algorithm execution times with
sufficient accuracy to single out the fastest algorithms.  This
micro-benchmark-based prediction proves to be several orders of magnitude faster
than executions of the actual algorithms.

\subsubsection{Tensor Notation.}
In the following, we denote tensor contractions by means of the Einstein
notation;\footnote{%
    For the sake of simplicity and without any loss of generality, we ignore any
    distinction between covariant and contravariant vectors; this means we treat
    any index as a subscript.
}
let us briefly explain said notation by means of an example. In the contraction
$C_{abc} = A_{ai} B_{ibc}$, the entries {\tt$C$[a,b,c]} of the resulting
three-dimensional tensor $C \in \mathbb R^{a \times b \times c}$ are computed as
$\forall \texttt a \forall \texttt b \forall \texttt c.  C\texttt{[a,b,c]} =
\sum_\texttt i A\texttt{[a,i]} B\texttt{[i,b,c]}$.  (In this notation, a
matrix-matrix product is denoted by $C_{ab} = A_{ai} B_{ib}$.) The indices that
appear in both tensors $A$ and $B$ --- the summation indices $i, j, \ldots$ ---
are called \emph{contracted}, while those that only appear in either $A$ or $B$
(and thus in $C$) --- $a, b, c, \ldots$ --- are called \emph{free} or
\emph{uncontracted}.  W.l.o.g., we assume that tensors are stored as
Fortran-style contiguous multidimensional arrays: vectors (1D tensors) are
stored contiguously, matrices (2D tensors) are stored as sequence of column
vectors, 3D tensors (visualized as cubes) are stored as a sequence of matrices
(planes of the cube), and so on.

\subsubsection{Related Work.}
The most prominent project targeting the efficient computation of tensor
contractions is probably the Tensor Contraction Engine, a compiler built
specifically for multi-tensor multi-index contractions to be executed within
memory constraints~\cite{tce}; in light of the  wide diffusion and nearly
optimal efficiency of the BLAS library, an extension to TCE was proposed to
compute contractions via BLAS operations~\cite{lu}.
In the same spirit, we provided simple rules to build a taxonomy for all
contractions between two tensors, identifying which BLAS routines are usable and
how to best exploit them~\cite{DiNapoli2014:210}.

\parsum{(performance prediction)}
There also exists a variety of work in the field of performance prediction in
the context of dense linear algebra. A notable example is
Iakymchuk~et~al.~\cite{roman,roman2}, where the authors model the performance of
dense linear algebra algorithms analytically based on very detailed models of
the occurring cache-misses.
Also, in~\cite{modeling}, we use measurement-based performance models to predict
the behavior of blocked algorithms.
However, none of these works target or address high-performance tensor
contractions and their peculiarities, i.e., very regular patterns in routine
invocation and memory access, but highly skewed dimensionality (tiny sizes for
at least one of the dimensions).

\subsubsection{Structure of the Paper.}
The rest of this paper is structured as follows.  The systematic generation of
BLAS-based algorithms for tensor contractions is discussed in
\autoref{sec:alggen}.  Our performance prediction framework is introduced in
\autoref{sec:pred}, and experimental results for a range of contractions are
presented \autoref{sec:results}.


    \section{Algorithm Generation}
    \label{sec:alggen}
    In this section, we briefly explain how to systematically generate a family of
BLAS-based algorithms for a tensor contraction.  For a detailed discussion of
the topic, we refer the reader to~\cite{DiNapoli2014:210}. 

Aware of the extreme level of efficiency inherent to the best BLAS
implementations, our approach for computing a contraction consists in reducing
it to a sequence of calls to one of the BLAS kernels.  Since BLAS operates on
scalars, vectors and matrices (zero-, one- and two-dimensional objects),
tensors must be expressed in terms of a collection of such objects.  To this
end, we introduce the concept of {\em slicing}: With the help of Matlab's
``{\tt :}'' notation,\footnote{%
  In the remainder of the paper, we use a Matlab-like notation: {\tt 1:$b$} are
  the numbers from 1 to $b$; an index {\tt :} in a tensor refers to all
  elements along that dimension, e.g., {\tt$C$[:,b]} is the $b$-th column of
  $C$.
} 
slicing a $d$-dimensional operand
$\mathcal Op \in \mathbb R^{n_1 \times n_2 \times \cdots \times n_d}$
along the $i$-th index (or dimension) means creating the $n_i$
$(d\!-\!1)$-dimensional slices
{\tt$\mathcal Op$[$\underbrace{\text{:,}\ldots\text{,:}}_{i-1}\text{,k,}\underbrace{\text{:,}\ldots\text{,:}}_{d-i}$]},
where $\texttt k = 1, \ldots, n_i$.

\begin{example}
Consider the matrix-matrix product 
$
  C_{ab} \coloneqq A_{ai} B_{ib}.
$
The slicing of the matrix $B$ along the $b$ dimension reduces the matrix to a
collection of column vectors; accordingly, the matrix-matrix product is reduced
to a sequence of matrix-vector operations:\footnote{%
    The pictogram next to the algorithm visualizes the slicing of the three
    tensors that originates a sequence of \gemv's.  The red objects represent
    the operands of the BLAS kernel.
}
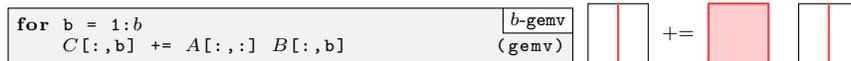
\begin{center}
    \vspace{-.5\baselineskip}\scriptsize
    \begin{algorithm}{b}{gemv}
        \begin{lstlisting}
for b = 1:$b$
    $C$[:,b] += $A$[:,:] $B$[:,b]!\hfill!(gemv)
        \end{lstlisting}
    \end{algorithm}
    \begin{tikzpicture}[scale=.8]
        \begin{drawsquare}
            \sliceB
        \end{drawsquare}
        \node at (1,0,0) {$\pluseq$};
        \begin{drawsquare}[(2,0,0)]
            \slicenone
        \end{drawsquare}
        \begin{drawsquare}[(3.5,0,0)]
            \sliceB
        \end{drawsquare}
    \end{tikzpicture}
\end{center}
Similarly, a multi-dimensional tensor contraction can be reduced to operations
involving solely matrices and vectors.
\end{example}

Depending on the slicing choices, a contraction is reduced to a number of
nested loops with one of the following kernels at the innermost loop's body:
\begin{itemize}
    \item BLAS-1:
        \begin{itemize}
            \item \dot (vector-vector inner product: $\alpha \coloneqq x^T y$),
            \item \axpy (vector scaling and addition: $y \coloneqq \alpha x + y$),
        \end{itemize}
    \item BLAS-2:
        \begin{itemize}
            \item \gemv (matrix-vector product: $y \coloneqq A x + y$),
            \item \ger (vector-vector outer product: $A \coloneqq x y^T + A$), and
        \end{itemize}
    \item BLAS-3:
        \begin{itemize}
            \item \gemm (matrix-matrix product: $C \coloneqq A B + C$).
        \end{itemize}
\end{itemize}
Notice that to comply with the BLAS interface, the elements in one of the two
dimensions of a matrix must be contiguous.  Therefore, algorithms that rely on
\gemv, \ger,, or \gemm as computational kernel may require a temporary copy of
slices prior and/or after the invocation of the corresponding BLAS routine.

As case study, let us consider the contraction
\begin{equation}
    C_{abc} = A_{ai} B_{ibc} \enspace ,
    \label{eq:case-study}
\end{equation}
which is visualized as follows:
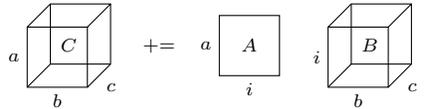
\begin{center}
    \scriptsize
    \begin{tikzpicture}[scale=.8]
        \begin{drawcube}
            \node[anchor=east] at (-1,0,1) {$a$};
            \node[anchor=north] at (0,-1,1) {$b$};
            \node[anchor=north west] at (1,-1,0) {$c$};
            \node {$C$};
        \end{drawcube}
        \node at (1.5,0,0) {$\pluseq$};
        \begin{drawsquare}[(3,0,0)]
            \node[anchor=east] at (-1,0,0) {$a$};
            \node[anchor=north] at (0,-1,0) {$i$};
            \node {$A$};
        \end{drawsquare}
        \begin{drawcube}[(5,0,0)]
            \node[anchor=east] at (-1,0,1) {$i$};
            \node[anchor=north] at (0,-1,1) {$b$};
            \node[anchor=north west] at (1,-1,0) {$c$};
            \node {$B$};
        \end{drawcube}
        \node at (6,0,0) {\vphantom{A}.};
    \end{tikzpicture}
\end{center}
Instead of a blind search for appropriate slicings, we generate algorithms by
following a goal-oriented approach: For each of the five kernels of interest,
we know the dimensionality required for each operand; accordingly, we deduce
how many slices are needed and which combination of free/contracted indices to
slice.  \autoref{tab:slicing-examples} (left) exhibits, for each kernel, the
conditions necessary for a contraction to be computed in terms of that kernel.
In particular, the second and the third columns indicate how many contracted
and free indices, respectively, appear in each kernel.  $A$ and $B$ refer to
the first and the second input operand of the kernel; in a contraction between
tensors of arbitrary dimension, all the indices beyond what indicated in these
columns must be sliced.

\begin{example}
    Since \gemm involves one free index in each of its operands $A$ and $B$,
    and one contracted index (common to both $A$ and $B$), in order to reduce a
    contraction to a sequence of \gemm calls, one must slice all free indices
    of $A$ but one, all free indices of $B$ but one, and all contracted indices
    but one.  With reference to~\eqref{eq:case-study}, this is achieved by
    slicing either dimension $b$ or $c$, resulting in the two algorithms
    (\algname{b}{gemm} and \algname{c}{gemm})\footnote{%
        The algorithm names are composed of two parts: the first part is the
        list of sliced tensor indices iterated over by the algorithm's loops
        and an apostrophe $'$ for each {\tt copy}-kernel, while the second part
        is the name of the used BLAS-kernel.
    }  shown in the last two examples of~\autoref{algs:ai_ibc}.
\end{example}

\begin{algorithms}
    \centering\scriptsize

    \begin{algorithm}{cab}{dot}
        \begin{lstlisting}
for c = 1:$c$
    for a = 1:$a$
        for b = 1:$b$
            $C$[a,b,c] += $A$[a,:] $B$[:,b,c]!\hfill!(dot)
        \end{lstlisting}
    \end{algorithm}
    \begin{tikzpicture}[scale=.8]
        \begin{drawcube}
            \sliceA[outer]
            \sliceB[outer]
            \sliceC[outer]
            \sliceAB[outer]
            \sliceAC[outer]
            \sliceBC[outer]
            \sliceABC
        \end{drawcube}
        \node at (1,0,0) {$\pluseq$};
        \begin{drawsquare}[(2,0,0)]
            \sliceA
        \end{drawsquare}
        \begin{drawcube}[(3.5,0,0)]
            \sliceB[outer]
            \sliceC[outer]
            \sliceBC
        \end{drawcube}
    \end{tikzpicture}

    \begin{algorithm}{bci}{axpy}
        \begin{lstlisting}
for b = 1:$b$
    for c = 1:$c$
        for i = 1:$i$
            $C$[:,b,c] += $A$[:,i] $B$[i,b,c]!\hfill!(axpy)
        \end{lstlisting}
    \end{algorithm}
    \begin{tikzpicture}[scale=.8]
        \begin{drawcube}
            \sliceB[outer]
            \sliceC[outer]
            \sliceBC
        \end{drawcube}
        \node at (1,0,0) {$\pluseq$};
        \begin{drawsquare}[(2,0,0)]
            \sliceB
        \end{drawsquare}
        \begin{drawcube}[(3.5,0,0)]
            \sliceA[outer]
            \sliceB[outer]
            \sliceC[outer]
            \sliceAB[outer]
            \sliceAC[outer]
            \sliceBC[outer]
            \sliceABC
        \end{drawcube}
    \end{tikzpicture}

    \begin{algorithm}{aib}{axpy}
        \begin{lstlisting}
for a = 1:$a$
    for i = 1:$i$
        for b = 1:$b$
            $C$[a,b,:] += $A$[a,i] $B$[i,b,:]!\hfill!(axpy)
        \end{lstlisting}
    \end{algorithm}
    \begin{tikzpicture}[scale=.8]
        \begin{drawcube}
            \sliceA[outer]
            \sliceB[outer]
            \sliceAB
        \end{drawcube}
        \node at (1,0,0) {$\pluseq$};
        \begin{drawsquare}[(2,0,0)]
            \sliceA[outer]
            \sliceB[outer]
            \sliceAB
        \end{drawsquare}
        \begin{drawcube}[(3.5,0,0)]
            \sliceA[outer]
            \sliceB[outer]
            \sliceAB
        \end{drawcube}
    \end{tikzpicture}

    \newcommand\contraction{ai_ibc}

    \begin{algorithm}{bc}{gemv}
        \begin{lstlisting}
for b = 1:$b$
    for c = 1:$c$
        $C$[:,b,c] += $A$[:,:] $B$[:,b,c]!\hfill!(gemv)
        \end{lstlisting}
    \end{algorithm}
    \begin{tikzpicture}[scale=.8]
        \begin{drawcube}
            \sliceB[outer]
            \sliceC[outer]
            \sliceBC
        \end{drawcube}
        \node at (1,0,0) {$\pluseq$};
        \begin{drawsquare}[(2,0,0)]
            \slicenone
        \end{drawsquare}
        \begin{drawcube}[(3.5,0,0)]
            \sliceB[outer]
            \sliceC[outer]
            \sliceBC
        \end{drawcube}
    \end{tikzpicture}

    \begin{algorithm}{ca}{gemv}
        \begin{lstlisting}
for c = 1:$c$
    for a = 1:$a$
        $C$[a,:,c] += $A$[a,:] $B$[:,:,c]!\hfill!(gemv)
        \end{lstlisting}
    \end{algorithm}
    \begin{tikzpicture}[scale=.8]
        \begin{drawcube}
            \sliceA[outer]
            \sliceC[outer]
            \sliceAC
        \end{drawcube}
        \node at (1,0,0) {$\pluseq$};
        \begin{drawsquare}[(2,0,0)]
            \sliceA
        \end{drawsquare}
        \begin{drawcube}[(3.5,0,0)]
            \sliceC
        \end{drawcube}
    \end{tikzpicture}

    \begin{algorithm}{ci}{ger}
        \begin{lstlisting}
for c = 1:$c$
    for i = 1:$i$
        $C$[:,:,c] += $A$[:,i] $B$[i,:,c]!\hfill!(ger)
        \end{lstlisting}
    \end{algorithm}
    \begin{tikzpicture}[scale=.8]
        \begin{drawcube}
            \sliceC
        \end{drawcube}
        \node at (1,0,0) {$\pluseq$};
        \begin{drawsquare}[(2,0,0)]
            \sliceB
        \end{drawsquare}
        \begin{drawcube}[(3.5,0,0)]
            \sliceA[outer]
            \sliceC[outer]
            \sliceAC
        \end{drawcube}
    \end{tikzpicture}

    \begin{algorithm}{bi}{ger}
        \begin{lstlisting}
for b = 1:$b$
    for i = 1:$i$
        $C$[:,b,:] += $A$[:,i] $B$[i,b,:]$^T$!\hfill!(ger)
        \end{lstlisting}
    \end{algorithm}
    \begin{tikzpicture}[scale=.8]
        \begin{drawcube}
            \sliceB
        \end{drawcube}
        \node at (1,0,0) {$\pluseq$};
        \begin{drawsquare}[(2,0,0)]
            \sliceB
        \end{drawsquare}
        \begin{drawcube}[(3.5,0,0)]
            \sliceA[outer]
            \sliceB[outer]
            \sliceAB
        \end{drawcube}
    \end{tikzpicture}

    \begin{algorithm}{c}{gemm}
        \begin{lstlisting}
for c = 1:$c$
    $C$[:,:,c] += $A$[:,:] $B$[:,:,c]!\hfill!(gemm)
        \end{lstlisting}
    \end{algorithm}
    \begin{tikzpicture}[scale=.8]
        \begin{drawcube}
            \sliceC
        \end{drawcube}
        \node at (1,0,0) {$\pluseq$};
        \begin{drawsquare}[(2,0,0)]
            \slicenone
        \end{drawsquare}
        \begin{drawcube}[(3.5,0,0)]
            \sliceC
        \end{drawcube}
    \end{tikzpicture}

    \begin{algorithm}{b}{gemm}
        \begin{lstlisting}
for b = 1:$b$
    $C$[:,b,:] += $A$[:,:] $B$[:,b,:]!\hfill!(gemm)
        \end{lstlisting}
    \end{algorithm}
    \begin{tikzpicture}[scale=.8]
        \begin{drawcube}
            \sliceB
        \end{drawcube}
        \node at (1,0,0) {$\pluseq$};
        \begin{drawsquare}[(2,0,0)]
            \slicenone
        \end{drawsquare}
        \begin{drawcube}[(3.5,0,0)]
            \sliceB
        \end{drawcube}
    \end{tikzpicture}
    
    \caption{
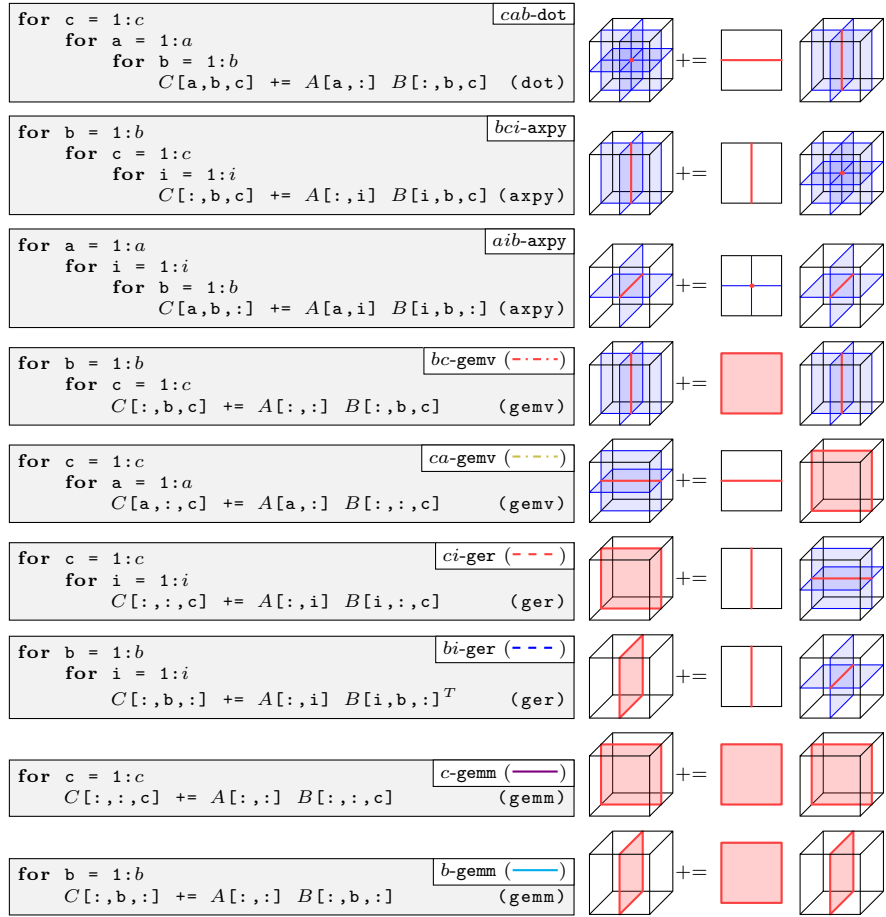
        $C_{abc} = A_{ai} B_{ibc}$: 9 exemplary algorithms out of
        36.\protect\footnotemark
    }
    \label{algs:ai_ibc}
\end{algorithms}
\footnotetext{%
    For algorithms with more than 1 {\tt for}-loop, all slicings are visualized
    in blue and only the kernel operands (the slicings' intersections) are in
    red.
}

\begin{table}[t]
    \centering
    \caption{Rules for tensor slicing to obtain a given BLAS kernel.  Left: how
        many contracted and how many free indices appear in the operation
        corresponding to a kernel.  Right: different slicings make it possible
        to express one contraction in terms of different kernels.  The names in
        the rightmost column refer to the algorithms in~\autoref{algs:ai_ibc}.
    }
    \label{tab:slicing-examples}
    \setlength{\tabcolsep}{6pt}
    \begin{tabular}{l|ll||l@{\hspace*{18pt}}l@{\hspace*{18pt}}l}
        \toprule
        Kernel  &\multicolumn{2}{c||}{Number of indices}    &\multicolumn{3}{c}{Examples from $C_{abc} = A_{ai} B_{ibc}$}\\\midrule
                &\multicolumn{1}{l}{contracted}             &\multicolumn{1}{c||}{free\quad\quad}   &kernel  &sliced  &resulting\\
                &                                           &                                       &indices &indices &algorithm\\
        \midrule
        \dot    &1  &0                                          &$i$        &$c,a,b$    &\algname{cab}{dot}\\[2pt]
        \multirow{2}{*}{\axpy}
                &\multirow{2}{*}{0}
                    &(1 in \!$A$ $\wedge$ 0 in \!$B)$ $\vee$    &$a$        &$b,c,i$    &\algname{bci}{axpy}\\
                &   &(0 in \!$A$ $\wedge$ 1 in \!$B)$           &$c$        &$a,i,b$    &\algname{aib}{axpy}\\[2pt]
        \multirow{2}{*}{\gemv}
                &\multirow{2}{*}{1}
                    &(1 in \!$A$ $\wedge$ 0 in \!$B)$ $\vee$    &$i, a$     &$b,c$      &\algname{bc}{gemv}\\
                &   &(0 in \!$A$ $\wedge$ 1 in \!$B)$           &$i, b$     &$c,a$      &\algname{ca}{gemv}\\[2pt]
        \ger    &0  &1 in $A$ $\wedge$ 1 in $B$                 &$a, c$     &$i, b$     &\algname{ib}{ger}\\[2pt]
        \gemm   &1  &1 in $A$ $\wedge$ 1 in $B$                 &$i, a, b$  &$c$        &\algname{c}{gemm}\\
        \bottomrule
    \end{tabular}
\end{table}

As already mentioned, given a contraction, there is no obvious a-priori choice
of kernel and slicings to attain the highest performance.  We therefore
generate all possible combinations.  Moreover, due to their impact on
performance and to further stress our modeling tool, we generate all the
permutations of the loops.

We developed a small algorithm and code generator that produces all such
algorithms, constructs for each of them a C-implementation, as well as an
abstract syntax tree (AST) representing its loop-based structure.  The ASTs are
then passed to the prediction tool introduced in the following section.


    \section{Performance Prediction}
    \label{sec:pred}
    \parsum{predictions from micro-benchmarks}
In this section, we present how to accurately model the performance of
algorithms that compute tensor contractions through BLAS kernels.  These
algorithms consist of one or more nested loops and cast all the computation in
terms of one single BLAS kernel. Taking advantage of this structure, we aim at
estimating the execution time of a target algorithm with the help of only few
micro-benchmarks of the kernels and with no direct execution of the algorithm
itself.  In order to obtain reliable estimates, the micro-benchmarks need to be
executed in a setup that mirrors as closely as possible the computing
environment (most importantly the cache) within the contraction algorithm. In
the following, we incrementally go through the steps required to build a
meaningful ``replica'' of the computing environment.

\parsum{contraction and setup}
Throughout this section, we track the changes in the performance prediction by
considering the exemplary contraction $C_{abc} = A_{ai} B_{ibc}$.  We chose the
tensors $A$ and $B$ of size $i = 8$ and $a = b = c = 8, \ldots, 1024$ --- a
deliberately challenging scenario due to the thin tensor dimension $i$, for
which BLAS kernels are generally not optimized.  Our generator produces 36
algorithms for the considered contraction, some of which are shown in
\autoref{algs:ai_ibc}:
\begin{itemize}
    \item 6 \dot-based,
    \item 18 \axpy-based,
    \item 6 \gemv-based: {
            \algname{bc}{gemv}~(\ref*{plt:ai_ibc:bc_gemv}),
            \algname{cb}{gemv}~(\ref*{plt:ai_ibc:cb_gemv}),
            \algname{ac}{gemv}~(\ref*{plt:ai_ibc:ac_gemv}),
            \algname{ca}{gemv}~(\ref*{plt:ai_ibc:ca_gemv}),
            \algname{ab}{gemv}~(\ref*{plt:ai_ibc:ab_gemv}),
            \algname{ba}{gemv}~(\ref*{plt:ai_ibc:ba_gemv})%
        },
    \item 4 \ger-based: {
            \algname{ci}{ger}~(\ref*{plt:ai_ibc:ci_ger}),
            \algname{ic}{ger}~(\ref*{plt:ai_ibc:ic_ger}),
            \algname{bi}{ger}~(\ref*{plt:ai_ibc:bi_ger}),
            \algname{ib}{ger}~(\ref*{plt:ai_ibc:ib_ger})%
        }, and
    \item 2 \gemm-based: {
            \algname{c}{gemm}~(\ref*{plt:ai_ibc:c_gemm}),
            \algname{b}{gemm}~(\ref*{plt:ai_ibc:b_gemm})%
        }.
\end{itemize}
In this section, to focus our attention, we will only consider the BLAS-2 and
BLAS-3 based algorithms (i.e., with kernels \gemv, \ger, and \gemm).

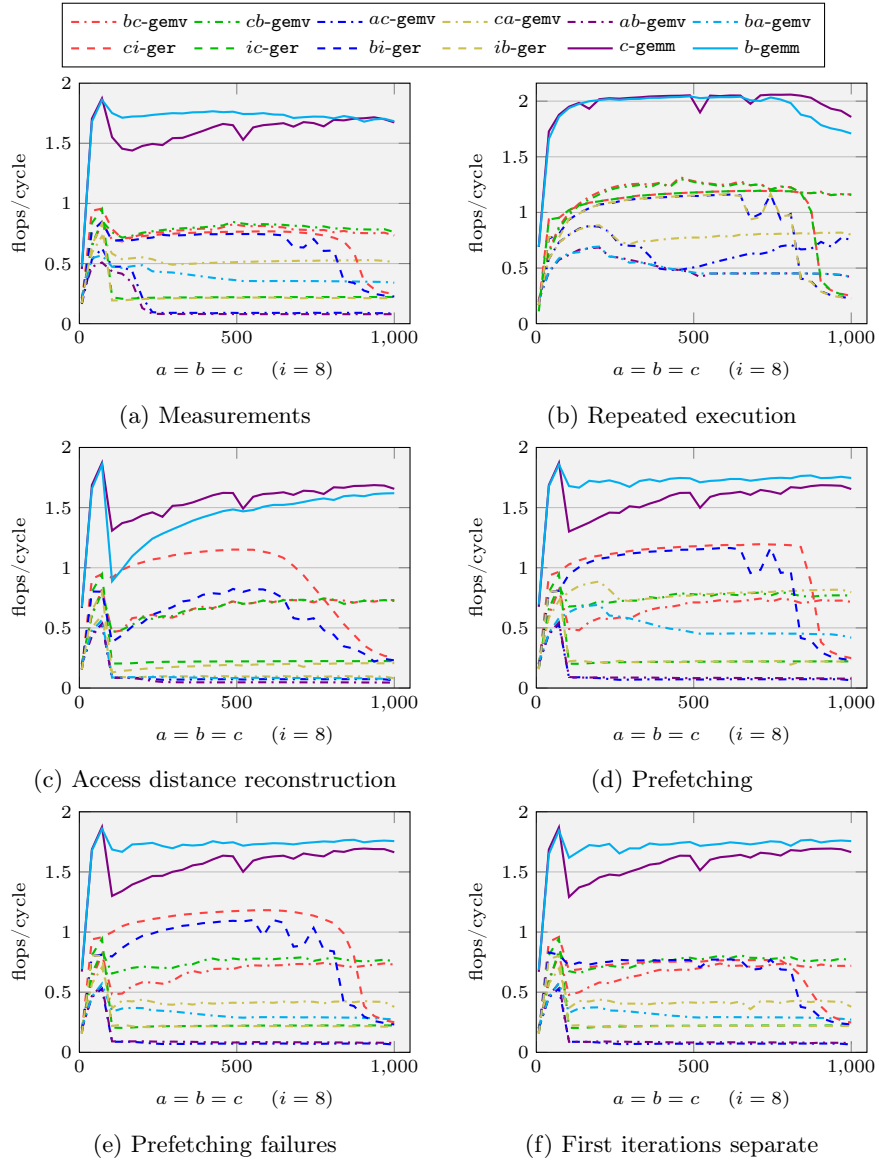
\begin{figure*}[p]
    \centering \scriptsize

    \ref*{leg:pred}

    \tikzset{external/export=true}

    \begin{subfigure}[b]{.49\textwidth}
        \centering
        \tikzsetnextfilename{pred_meas}
        \begin{tikzpicture}
            \begin{axis}[predplt,
                    twocolplot,
                    legend to name=leg:pred,
                    legend columns=6,
                    ymax=2,
                ]
                \foreach \var/\loops/\kernel in {%
                        25/bc/gemv,%
                        26/cb/gemv,%
                        27/ac/gemv,%
                        28/ca/gemv,%
                        29/ab/gemv,%
                        30/ba/gemv,%
                        31/ci/ger,%
                        32/ic/ger,%
                        33/bi/ger,%
                        34/ib/ger,%
                        35/c/gemm,%
                        36/b/gemm%
                    } {
                    \plot file {figures/data/pred/meas/var\var.min};
                    \addlegendentryexpanded{\algname{\loops}{\kernel}}
                    \label{plt:ai_ibc:\loops_\kernel}
                }
            \end{axis}
        \end{tikzpicture}
        \caption{Measurements}
        \label{fig:pred_meas}
    \end{subfigure}
    \begin{subfigure}[b]{.49\textwidth}
        \centering
        \tikzsetnextfilename{pred_step}
        \begin{tikzpicture}
            \begin{axis}[predplt, twocolplot]
                \foreach \var in {25,...,36}
                    \plot file {figures/data/pred/step/var\var.min};
            \end{axis}
        \end{tikzpicture}
        \caption{Repeated execution}
        \label{fig:pred_step}
    \end{subfigure}

    \begin{subfigure}[b]{.49\textwidth}
        \centering
        \tikzsetnextfilename{pred_step+s}
        \begin{tikzpicture}
            \begin{axis}[predplt, ymax=2, twocolplot]
                \foreach \var in {25,...,36}
                    \plot file {figures/data/pred/step+s/var\var.min};
            \end{axis}
        \end{tikzpicture}
        \caption{Access distance reconstruction}
        \label{fig:pred_step+s}
    \end{subfigure}
    \begin{subfigure}[b]{.49\textwidth}
        \centering
        \tikzsetnextfilename{pred_step+sp}
        \begin{tikzpicture}
            \begin{axis}[predplt, ymax=2, twocolplot]
                \foreach \var in {25,...,36}
                    \plot file {figures/data/pred/step+sp/var\var.min};
            \end{axis}
        \end{tikzpicture}
        \caption{Prefetching}
        \label{fig:pred_step+sp}
    \end{subfigure}

    \begin{subfigure}[b]{.49\textwidth}
        \centering
        \tikzsetnextfilename{pred_step+spx}
        \begin{tikzpicture}
            \begin{axis}[predplt, ymax=2, twocolplot]
                \foreach \var in {25,...,36}
                    \plot file {figures/data/pred/step+spx/var\var.min};
            \end{axis}
        \end{tikzpicture}
        \caption{Prefetching failures}
        \label{fig:pred_step+spx}
    \end{subfigure}
    \begin{subfigure}[b]{.49\textwidth}
        \centering
        \tikzsetnextfilename{pred_pred}
        \begin{tikzpicture}
            \begin{axis}[predplt, ymax=2, twocolplot]
                \foreach \var in {25,...,36}
                    \plot file {figures/data/pred/pred/var\var.min};
            \end{axis}
        \end{tikzpicture}
        \caption{First iterations separate}
        \label{fig:pred_pred}
    \end{subfigure}

    \tikzset{external/export=false}
    \caption{
        $C_{abc} \coloneqq A_{ai} B_{ibc}$: Performance measurements and
        various stages of performance predictions (BLAS-2 and BLAS-3).
    }
    \label{fig:pred}
\end{figure*}

We execute these algorithms on 1 core of an Intel Harpertown E5450
CPU\footnote{%
    2 GHz, 4 cores, 4 double precision flops/cycle/core, 6MB L2 cache/2 cores.
} linking with the {\sc OpenBLAS} library~\cite{openblas}.
\hyperref[fig:pred_meas]{Figure~\ref*{fig:pred_meas}} displays the performance,
in terms of computed floating point operations per clock cycle (flops/cycle),
measured for each algorithm; our goal is to accurately reproduce, without
executing the algorithms, such performance profiles.  While it is evident that
only two of the algorithms --- the \gemm-based
\algname{c}{gemm}~(\ref*{plt:ai_ibc:c_gemm}) and
\algname{b}{gemm}~(\ref*{plt:ai_ibc:b_gemm}) --- are competitive, we aim at
predicting the behavior of all the algorithms to demonstrate the broad
applicability of our methodology.

\subsection{Repeated Execution}
\parsum{starting point: repeated execution}
The first, most intuitive, attempt to predict the performance of an algorithm
relies on the isolated and repeated measurement of its BLAS kernel.  We
implemented this approach by executing each kernel ten times and extracting the
median execution time; the corresponding estimate is then obtained by
multiplying the median by the number of kernel invocations within the algorithm.
In our example, this boils down to multiplying the kernel execution time with
the product of all loop lengths.

\parsum{insufficient accuracy <= assumed all in cache}
The performance profiles predicted by this first, rough approach are shown in
\autoref{fig:pred_step}. By comparing this figure with the reference
\autoref{fig:pred_meas}, it becomes apparent that while the two top algorithms
are already correctly identified, the performance of almost all algorithms is
consistently overestimated.  In other words, when executed as part of the
algorithms, the BLAS kernels take longer to complete than in the isolated
micro-benchmarks.  The reason for this discrepancy is that the micro-benchmarks
invoke the kernels repeatedly, with the same memory regions as operands, i.e.,
they operate on warm data (the operands remain in the CPU's cache).  Within the
algorithm, by contrast, at least one, and potentially even all of the operands,
vary from one invocation to the next, i.e., the kernels operate at least
partially on cold data.

\subsection{Operand Access Distance}
\parsum{access distance}
In order to improve the accuracy of the predictions, the idea is to first
identify the status of the cache in the algorithm prior to the invocation of the
BLAS kernel (``precondition''), and then to replicate such a status in the
micro-benchmark.  For this purpose, each algorithm is symbolically analyzed to
reconstruct the order of memory accesses involving the kernel's operands.  For
each operand, we determine the set of memory regions $M$ that were loaded into
cache since its last access, and define the {\em access distance} as the sum of
the size of these regions $M$.

\parsum{cache setup}
Once the access distances for all operands of a kernel are determined, we can
create an artificial sequence of memory accesses to reconstruct the cache
precondition.  Based on this cache setup, the BLAS kernels are timed in a
micro-benchmark that closely resembles the actual execution of the algorithm.
As before, these micro-benchmarks are repeated and timed ten times to yield a
stable median.  From the median, the performance of the algorithms is again
obtained based on the number of kernel invocations per algorithm execution.

\parsum{assumption}
To predict which memory regions are in cache, we assume a fully associative
Least Recently Used (LRU) cache replacement policy\footnote{%
  Due to the regular storage format and memory access strides of dense linear
  algebra operations such as the considered tensor contractions, this
  simplifying assumption does not affect the reliability of the results.
} and sum up the size of all memory regions accessed since an operand's last
use, yielding the access distance.  In first instance, we also assume that all
loops surrounding the kernel are somewhere in the middle of their traversal
(i.e., not in their first iteration); this assumption will be lifted later.

\parsum{one kernel only case}
We now describe how to obtain the access distance for each of the operands.  The
presented method is general and allows for any combinations of loops and
multiple kernels within the abstract syntax tree (AST), however for the sake of
clarity, we limit the discussion to ASTs that only consist of a series of loops
with a single call to a BLAS kernel at their innermost loop.

\parsum{back-traversal through AST}
For each operand $\mathcal Op$, we examine the algorithm's AST (see
\autoref{sec:alggen}) with the kernel of interest as a starting point.  The AST
is traversed backwards until the previous access to $\mathcal Op$ (or the AST's
root) is found, thereby collecting all other operands involved in kernels in the
initially empty set $M$.  Going up the AST, three different cases can be
encountered.
\begin{enumerate}
    \item {\bf $\mathcal{O}p$ does not vary across the surrounding loop.}

        \begin{example}
            \sloppy
            In algorithm \algname{ca}{gemv}~(\ref*{plt:ai_ibc:ca_gemv}),
            repeated below, the operand {\tt$B$[:,:,c]} does not depend on the
            surrounding loop's iterator {\tt a}. Hence, $M = \emptyset$ and the
            operand's access distance is 0.
            
            \vspace{.5ex}
            \centering\scriptsize
            \newcommand\contraction{ai_ibc}
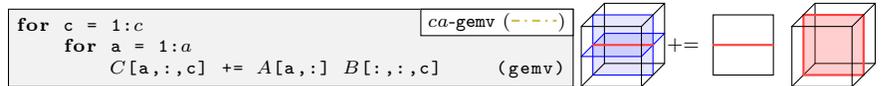
            \begin{algorithm}{ca}{gemv}
                \begin{lstlisting}
for c = 1:$c$
    for a = 1:$a$
        $C$[a,:,c] += $A$[a,:] $B$[:,:,c]!\hfill!(gemv)
                \end{lstlisting}
            \end{algorithm}%
            \begin{tikzpicture}[scale=.8]
                \begin{drawcube}
                    \sliceA[outer]
                    \sliceC[outer]
                    \sliceAC
                \end{drawcube}
                \node at (1,0,0) {$\pluseq$};
                \begin{drawsquare}[(2,0,0)]
                    \sliceA
                \end{drawsquare}
                \begin{drawcube}[(3.5,0,0)]
                    \sliceC
                \end{drawcube}
            \end{tikzpicture}
        \end{example}

        $\mathcal{O}p$ refers to the same memory region as in the previous
        iteration of the surrounding loop.  The back-traversal therefore
        terminates and the memory regions collected in $M$ so far determine the
        access distance.

    \item {\bf $\mathcal Op$ varies across the surrounding loop.}

        \begin{example}
            In algorithm \algname{ca}{gemv}~(\ref*{plt:ai_ibc:ca_gemv}), the
            operand {\tt$A$[a,:]} depends on the surrounding loop's iterator
            {\tt a}.
        \end{example}

        $\mathcal Op$ referred to a different memory region in the previous
        iteration of the loop.  As a result, it is safe to assume that at least
        all memory regions covered by all kernel operands throughout these loops
        were accessed since the last access to $\mathcal Op$.  Hence, all
        operands are added to $M$ and the memory regions are symbolically joined
        along the dimensions the loop iterated over.  

        \addtocounter{example}{-1}
        \begin{example}[continued]
            \sloppy
            The algorithm's kernel operates on {\tt$A$[a,:]}, {\tt$B$[:,:,c]},
            and {\tt$C$[a,:,c]}. Joining these operands across the index {\tt a}
            yields the memory regions {\tt$M = \{A\text{[:,:]}, B\text{[:,:,c]},
            C\text{[:,:,c]}\}$}.
        \end{example}

        Since a previous access to $\mathcal Op$ was not yet detected, the
        traversal proceeds by going up one level in the AST, and applying the
        method recursively: the surrounding loop now takes the role of the
        starting node and we look for a previous access $\mathcal Op$ joined
        across this loop.

        \addtocounter{example}{-1}
        \begin{example}[continued]
            The back-traversal now looks for a previous access to {\tt$A$[:,:]}
            ({\tt$A$[a,:]} joint across {\tt a}) on the second-innermost loop.
            This time, the region is independent of the surrounding loop's
            iterator {\tt c}; therefore, in this second step, case 1. above
            applies and the access distance is computed from the previously
            collected set {\tt $M = \{A\text{[:,:]}, B\text{[:,:,c]},
            C\text{[:,:,c]}\}$}.
        \end{example}
    \item {\bf The parent node is the AST's root.}

        \begin{example}
            In algorithm \algname{ca}{gemv}~(\ref*{plt:ai_ibc:ca_gemv}), the
            operand {\tt$C$[a,:,c]} depends on both of the surrounding loops'
            iterators {\tt a} and {\tt c}.  Therefore, the back-traversal
            encounters case 2. above in both its first and second step, joining
            the kernel's operands {\tt$A$[a,:]}, {\tt$B$[:,:,c]}, and
            {\tt$C$[a,:,c]} across first {\tt a} and then {\tt c}, yielding
            {\tt$M = \{A\text{[:,:]}, B\text{[:,:,:]}, C\text{[:,:,:]}\}$}.  In
            the third step of the back-traversal, the outermost loop is already
            the starting point --- the algorithm's root is reached.
        \end{example}

        In this case, the considered region is accessed only once (and for the
        first time).  Since we do not know how the contraction is used (within a
        surrounding program), we can generally not make any assertions on the
        access distance.  For the purpose of this paper, in which we execute the
        contraction repeatedly to measure its performance, however, we assume
        that no further memory regions were loaded since the last invocation of
        the contraction --- i.e., we compute the access distance from the
        previously collected memory regions in $M$.
\end{enumerate}

\parsum{generate setup}
Based on the such obtained access distances for each operand of an algorithm's
kernel, we now construct a list of memory accesses that emulates the accesses
within the algorithm prior to the kernel's execution.  This list consists of
accesses to the kernel's operands, interleaved with accesses to remote memory
regions, in order to flush portions of the cache corresponding to the access
distances:  First, we access the operand with the largest access distance, then
a remote region that accounts for the difference to the next smaller access
distance, followed by the next operand, and so on until the operands with the
smallest access distance followed by a remote access of this size.  If the
access distances to the first operands in this list are larger than $\frac54$
times the cache size, the list is truncated down to this limit at the front.

\begin{example}
    For algorithm \algname{ca}{gemv}~(\ref*{plt:ai_ibc:ca_gemv}), the following
    table summarizes the operands $O$, their sizes $s$, the corresponding
    collections $M$ and the implicated access distances $d$ for contraction
    sizes $a = b = c = 400$ and $i = 8$ (all sizes in doubles = 8 bytes):
    \begin{center}
        \small
        \vspace{-1ex}
        \setlength{\tabcolsep}{1em}
        \begin{tabular}{lrlr}
            \toprule
            $O$ &$s$ &$M$ &$d$ \\
            \midrule
            {\tt$B$[:,:,c]} &$3{,}200$  &$\emptyset$                                                &$0$\\
            {\tt$A$[a,:]}   &$8$        &{\tt$\{A\text{[:,:]}, B\text{[:,:,c]}, C\text{[:,:,c]}\}$} &$166{,}400$\\
            {\tt$C$[a,:,c]} &$400$      &{\tt$\{A\text{[:,:]}, B\text{[:,:,:]}, C\text{[:,:,:]}\}$} &$65{,}283{,}200$\\
            \bottomrule
        \end{tabular}
    \end{center}

    From these distances, we get the following list of memory accesses as a
    setup for the \gemv-kernel, where $[s]$ correspond to remote memory accesses
    of size $s$:
    \begin{center}
        {\tt$C$[a,:,c]},
        $[65{,}116{,}792]$,
        {\tt$A$[a,:]},
        $[163{,}200]$,
        {\tt$B$[:,:,c]}.
    \end{center}
    Note, that remote accesses are not directly of the access distance's sizes;
    however, this size is reached for each operand as the sum of the sizes of
    all accesses to its right in this list. (e.g., the access distances of
    {\tt$A$[a,:]} is reached as {\tt$163{,}200 +
    \mathrm{sizeof}(B\text{[:,:,c]}) = 166{,}400$}).

    Now, the largest access distance is at $65{,}283{,}200$ considerably larger
    than $983{,}040$ ($\frac54$ times the cache size of $\frac{6\mathrm{MB}}8 =
    786{,}432$ doubles).  Hence, the list is cut at this size, yielding the
    final setup for this algorithm's micro-benchmark:
    \begin{center}
        $[816{,}632]$,
        {\tt$A$[a,:]},
        $[163{,}200]$,
        {\tt$B$[:,:,c]}.
    \end{center}
\end{example}

The thus obtained benchmark, consisting of the setup followed by the kernel
invocation, is once more executed ten times.  The median of the kernel run-times
of these ten benchmarks is then used to compute our second execution time
estimate.

\parsum{improvements and shortcomings of the new estimates}
In \autoref{fig:pred_step+s}, we present the flops/cycle performance of our new
estimates.  These predictions are much closer to the measured performance
(\autoref{fig:pred_meas}) than the first rough estimates
(\autoref{fig:pred_step}).  For several algorithms (such as
\algname{ic}{ger}~(\ref*{plt:ai_ibc:ic_ger}), \autoref{algs:ai_ibc}), the error
is already within a few percent; for many others instead, the predictions are
still off.  In particular, the performance of some algorithms --- for instance,
\algname{bi}{ger}~(\ref*{plt:ai_ibc:bi_ger}) (see \autoref{algs:ai_ibc}) --- is
underestimated; this is due to the fact that based on the access distance,
certain operands are placed out of cache, while in practice they are (partially)
brought into cache through either prefetching or because they share cache-lines
across loop iterations.  We address this discrepancy by further refining our
micro-benchmarks.

\subsection{Cache Prefetching}
\parsum{explanation prefetching}
In the considered type of tensor contraction algorithms, prefetching of operands
or sharing of cache lines across loop iterations occur frequently.

\begin{example}
    In algorithm \algname{bi}{ger}~(\ref*{plt:ai_ibc:bi_ger}), the operand
    {\tt$A$[:,i]} points to a different memory location in each iteration of the
    inner loop across {\tt i}.  However, these vectors-operands are consecutive
    in memory; thus, when reaching the end of {\tt$A$[:,i]}, the prefetcher will
    likely already load the next memory elements, which constitute {\tt$A$[:,i]}
    in the next iteration.  Likewise, operand {\tt$B$[i,b,:]} varies across
    inner loop iterations; however, since this loop iterates over the region's
    first dimension {\tt i}, 8 consecutive operands\footnote{%
        The cache-line size is $64\mathrm{B} = 8$ doubles.
    } {\tt$B$[i,b,:]} will occupy the same cache-line.
\end{example}

\parsum{detect prefetching}
Such prefetching situations occur when a certain set of conditions are met,
namely:
\begin{enumerate}
    \item the operand varies across the directly surrounding loop, and
    \item the iterator of this loop indexes
        \begin{itemize}
            \item either the first dimension of the operand, 
            \item or its second dimension, while the first is accessed entirely,
                or fits in a single cache-line.
        \end{itemize}
\end{enumerate}
As part of our AST-based algorithm analysis, such conditions are tested; when
both of them are met, we can use a slight modification of the previously
introduced method to compute the {\em prefetch distance}, i.e., how long ago the
prefetching occurred.  These prefetch distances are then integrated into the
micro-bench\-mark's setup list just like the access distances, only that for
prefetch accesses the access is limited to one cache-line along an operand's
first dimension.

\begin{example}
    In algorithm \algname{ca}{gemv}~(\ref*{plt:ai_ibc:ca_gemv}), for which we
    explicitly constructed the setup list in the previous section, both operands
    {\tt$A$[a,:]} and {\tt$C$[a,:,b]} meet both of the prefetching conditions:
    1. they vary across the surrounding loop iterator {\tt a} and 2. {\tt a}
    indexes their first dimensions (sharing of cache-lines).  As a result, their
    prefetch distances are $0$ and the prefetching access will access the entire
    operands since their extension along the first, contiguously stored
    dimension is $1$.  Since the remaining operand {\tt$B$[:,:,c]} has an access
    distance of $0$, all operands are now accessed immediately before the kernel
    invocation; the setup list is reduced to
    \begin{center}
        {\tt$C$[a,:,c]},
        {\tt$A$[a,:]},
        {\tt$B$[:,:,c]}.
    \end{center}
    (Since this setup consists only of accesses to the operands, it becomes
    redundant in out micro-benchmarks, because each of the ten repetitions will
    already touch all operands for the next repetition; hence, in such a case,
    we omit the setup altogether.)
\end{example}

\parsum{prefetching results}
Now accounting for prefetching, we obtain the performance estimates shown in
\autoref{fig:pred_step+sp}.  Here, several algorithms,
e.g.~\algname{ba}{gemv}~(\ref*{plt:ai_ibc:ba_gemv}), are estimated closer to
their measured performance.  However, several other algorithms, including
\algname{ca}{gemv}~(\ref*{plt:ai_ibc:ca_gemv}) are overestimated in performance
(i.e., underestimated in execution time).  There are two separate causes for
this discrepancy.
\begin{itemize}
    \item In several algorithms, such as
        \algname{ca}{gemv}~(\ref*{plt:ai_ibc:ca_gemv}), where prefetching
        implicitly happens due to sharing of cache-lines, the prefetcher fails
        once a new cache-line is reached.
    \item In other algorithms, such as
        \algname{bi}{ger}~(\ref*{plt:ai_ibc:bi_ger}), the innermost loop is so
        short (here: 8 iterations) that each first iteration of the loop
        significantly impacts performance.
\end{itemize}
These two causes are teated in separately in the following sections.

\subsection{Prefetching Failures}
\parsum{prefetching along first tensor dimension fails every 8}
For those algorithms in which certain operands are identified as prefetched
because they share cache lines across iterations (i.e., the surrounding loop
indexes their first dimension), the CPU would need to prefetch the next
cache-line every 8 iterations (1 cache-line = 8 doubles).  However, as a
detailed analysis of hand-instrumented algorithms has shown, the CPU fails to do
so.  As a result, in every 8th iteration of the innermost loop, the operand is
not available and the kernel may take significantly longer.

\parsum{new, separate benchmark}
We account for this prefetching-artifact by performing two separate
micro-benchmarks: one simulating the 7 iterations, in which the operand is
available in cache, as before, and one for the 8th iteration, where we account
for the failure to prefetch and eliminate the emulated prefetching from our
setup-list.  The prediction for the total execution time is now obtained from
weighting these two benchmark timings according to their number of occurrences
in the algorithm and summing them up.

\begin{example}
    In algorithm \algname{ca}{gemv}~(\ref*{plt:ai_ibc:ca_gemv}), the memory
    regions of both {\tt$A$[a,:]} and {\tt$C$[a,:,c]}, respectively, share
    cache-lines across iterations of the innermost loops over {\tt a}.  Hence,
    affecting not one but two of the kernel's operands, every 8th iteration the
    kernel execution time increases drastically by a factor of about $4.5$.  To
    account for these ``prefetching failures'', we introduce a second set of
    micro-benchmarks, where the prefetching emulating accesses are removed from
    the setup list, resulting for $a = b = c = 400$ and $i = 8$, as without
    prefetching, in:
    \begin{center}
        $[816{,}632]$,
        {\tt$A$[a,:]},
        $[163{,}200]$,
        {\tt$B$[:,:,c]}.
    \end{center}
\end{example}

\parsum{prefetching failure results}
\autoref{fig:pred_step+spx} shows the predictions obtained after this
improvement.  Most noticeably in \algname{ca}{gemv}~(\ref*{plt:ai_ibc:ca_gemv}),
the overestimation of algorithms where iterations share cache-lines are now
corrected.

\subsection{First Loop Iterations}
\parsum{first iteration way off}
The predictions for several algorithms, such as
\algname{ci}{ger}~(\ref*{plt:ai_ibc:ci_ger}), are still severely off, because
the innermost loop of these algorithms is very short (in our example 8
iterations long).  In such a case, the predictions are very accurate for all but
the first iteration.  Due to vastly different cache preconditions for this first
iteration, however, its performance can be significantly different (in our case,
up to $10\times$ slower).  Combined with the low total iteration count, this
results in predictions that are off by a factor of up to $2$.

\parsum{micro-benchmark first iteration separately}
To treat such situations, we introduce separate benchmarks to predict the
performance of the first iteration of the innermost loop (and further loops if
their first iterations account for more than 1\% of the total kernel
invocations).  For this purpose, the access distance evaluation method is
slightly modified: instead of the kernel itself, the starting point is now the
loop whose first iteration is considered, and the set $M$ already contains all
of the kernel's memory regions joined across this loop.

\begin{example}
    In algorithm \algname{ci}{ger}~(\ref*{plt:ai_ibc:ci_ger}), the innermost
    loop over $i$ is in our example only $8$ iterations long.  For all but the
    first iteration, the operand {\tt$C$[:,:,c]} stays the same, while
    {\tt$A$[:,i]} and {\tt$B$[i,:,c]} are prefetched, leading to optimal
    conditions for performance.  In the first iteration (i.e., the next {\tt c}
    iteration) however, {\tt$C$[:,:,c]} refers to a different memory location
    and prefetching fails for {\tt$A$[:,i]}, leading to severely lower
    performance.
\end{example}

\parsum{weight and sum separate timings}
From these improved access distances, the cache setup and micro-benchmark are
performed just as before.  As for the ``prefetching failures'', the prediction
for the total execution time is now obtained from weighting of all relevant
benchmark timings with the corresponding number of occurrences within the
algorithm.

\parsum{results}
In \autoref{fig:pred_pred}, we present the improved performance predictions
obtained from this modification.  The performance of all algorithms is now
predicted with satisfying accuracy.


    \section{Results}
    \label{sec:results}
    In order to showcase its applicability and effectiveness, in this section we
apply our technique for performance prediction to a range of contractions.  We
consider three test cases: In \autoref{sec:ai_ibc2}, we use different hard- and
software, as well as changing the problem sizes. In \autoref{sec:noblas3}, we
consider a contraction that only allows the use of BLAS-1 and BLAS-2.  Finally,
in \autoref{sec:ijb_jcid}, we consider a more complex contraction with numerous
alternative algorithms and multithreading.

\subsection{Test 1: $C_{abc} = A_{ai} B_{ibc}$, Different Setup}
\label{sec:ai_ibc2}
\begin{figure}[t]
    \centering \scriptsize

    \ref*{leg:pred}
    \ref*{leg:ai_ibc_other}

    \vspace{.5\baselineskip}

    \tikzset{external/export=true}

    \begin{subfigure}{.49\textwidth}
        \centering
        \tikzsetnextfilename{ai_ibc2_meas}
        \begin{tikzpicture}
            \begin{axis}[predplt,
                    twocolplot,
                    xlabel={$i$ \quad ($a = b = c = 128$)},
                    ymax=7,
                    legend to name=leg:ai_ibc_other,
                    legend columns=-1,
                ]
                \addlegendimage{plot7, dotted}
                \addlegendentry{{\tt dot}-based}
                \addlegendimage{plot8, dotted}
                \addlegendentry{{\tt axpy}-based}
                \foreach \var in {25,...,36}
                    \addplot file {figures/data/ai_ibc2/meas/var\var.min};
                \foreach \var in {1,...,6}
                    \addplot[plot7, dotted] file {figures/data/ai_ibc2/meas/var\var.min};
                \foreach \var in {7,...,24}
                    \addplot[plot8, dotted] file {figures/data/ai_ibc2/meas/var\var.min};
            \end{axis}
        \end{tikzpicture}
        \caption{measurements}
    \end{subfigure}
    \begin{subfigure}{.49\textwidth}
        \centering
        \tikzsetnextfilename{ai_ibc2_pred}
        \begin{tikzpicture}
            \begin{axis}[predplt,
                    twocolplot,
                    xlabel={$i$ \quad ($a = b = c = 128$)},
                    ymax=7
                ]
                \foreach \var in {25,...,36}
                    \addplot file {figures/data/ai_ibc2/pred/var\var.min};
                \foreach \var in {1,...,6}
                    \addplot[plot7, dotted] file {figures/data/ai_ibc2/pred/var\var.min};
                \foreach \var in {7,...,24}
                    \addplot[plot8, dotted] file {figures/data/ai_ibc2/pred/var\var.min};
            \end{axis}
        \end{tikzpicture}
        \caption{predictions}
    \end{subfigure}

    \tikzset{external/export=false}
    \caption[]{
        $C_{abc} \coloneqq A_{ai} B_{ibc}$: Performance measurements and predictions.
    }
    \label{fig:ai_ibc2}
\end{figure}
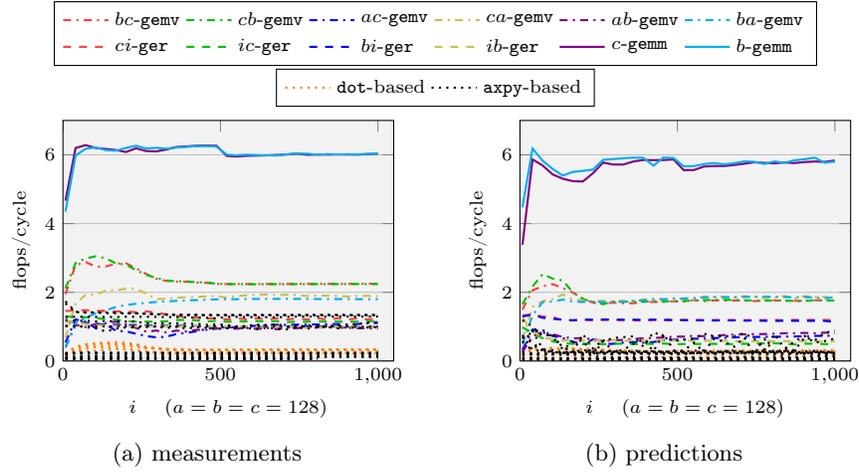

We commence with the same contraction used as case study in the previous
section, yet with an entirely different setup: the sizes of $a$, $b$, and $c$
are now fixed to 128, while the value of $i$ ranges from 8 to 1{,}024.  As
experimental environment, we use a 10-core Intel Ivy Bridge-EP E5-2680 v2
processor running at a frequency of 3.6 GHz (Turbo) and 25 MB of L3 cache. Each
core can execute 8 double precision flops/cycle.  The routines for both the
actual measurements and the micro-benchmarks were linked to the Intel Math
Kernel Library (MKL, version 11.0) BLAS implementation.
\hyperref[fig:ai_ibc2]{Figure~\ref*{fig:ai_ibc2}} 
contains the performance measurements and the corresponding predictions for all
36 algorithms (see \autoref{algs:ai_ibc}).  Although everything, ranging from
the problem size to the machine and BLAS library was changed in this setup, the
predictions are of equivalent quality and our tool correclty determines that the
\gemm-based algorithms (\algname{c}{gemm}~(\ref*{plt:ai_ibc:c_gemm}) and
\algname{b}{gemm}~(\ref*{plt:ai_ibc:b_gemm})) perform best and equally well.

\subsection{Test 2: $C_a = A_{iaj} B_{ji}$, only BLAS-1 and BLAS-2}
\label{sec:noblas3}
\begin{figure}[t]
    \centering \scriptsize

    \ref*{leg:iaj_ji}

    \tikzset{external/export=true}

    \begin{subfigure}{.49\textwidth}
        \tikzsetnextfilename{iaj_ji_meas}
        \begin{tikzpicture}
            \begin{axis}[perfplot,
                    twocolplot,
                    cycle list name=default,
                    legend to name=leg:iaj_ji,
                    legend columns=4,
                    ymax=1.5,
                    xlabel={$a = i = j$}
                ]
                \foreach \var/\loops/\kernel in {%
                        1/aj/dot,%
                        2/ja/dot,%
                        3/ai/dot,%
                        4/ia/dot,%
                        5/ij/axpy,%
                        6/ji/axpy,%
                        7/j/gemv,%
                        8/i'/gemv%
                }{
                    \addplot file {figures/data/iaj_ji/meas/var\var.min};
                    \addlegendentryexpanded{\algname{\loops}{\kernel}}
                    \label{plt:iaj_ji:\loops_\kernel}
                }
            \end{axis}
        \end{tikzpicture}
        \caption{measurements}
    \end{subfigure}
    \begin{subfigure}{.49\textwidth}
        \tikzsetnextfilename{iaj_ji_pred}
        \begin{tikzpicture}
            \begin{axis}[perfplot,
                    twocolplot,
                    cycle list name=default,
                    ymax=1.5,
                    xlabel={$a = i = j$}
                ]
                \foreach \var in {1,...,8}
                    \addplot file {figures/data/iaj_ji/pred/var\var.min};
            \end{axis}
        \end{tikzpicture}
        \caption{predictions}
    \end{subfigure}

    \tikzset{external/export=false}
    \caption[]{
        $C_{a} \coloneqq A_{iaj} B_{ji}$: Performance measurements and
        predictions.
    }
    \label{fig:iaj_ji}
\end{figure}
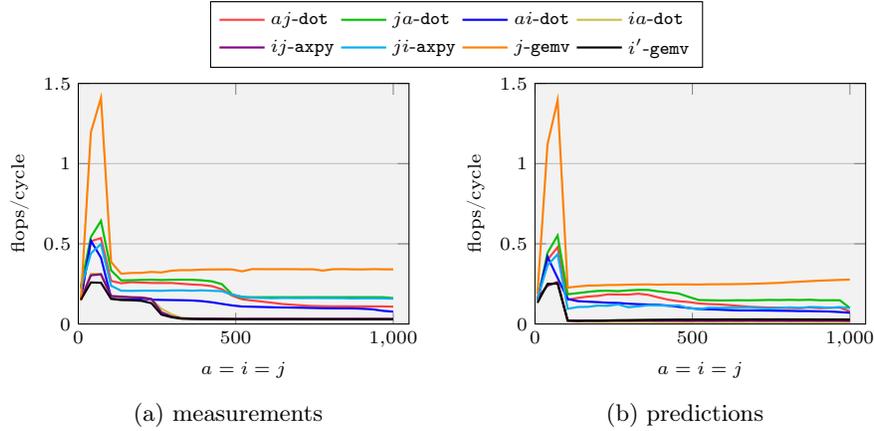

For certain contractions (e.g., those involving 1D tensors), \gemm cannot be
used as a compute kernel, and only algorithms based on BLAS-2 or BLAS-1 are
possible.  One such scenario is encontered in the contraction $C_a = A_{iaj}
B_{ji}$, for which our generator yields 8 BLAS-based algorithms:
\begin{itemize}
    \item 4 \dot-based: 
        \algname{aj}{dot}~(\ref*{plt:iaj_ji:aj_dot}),
        \algname{ja}{dot}~(\ref*{plt:iaj_ji:ja_dot}),
        \algname{ai}{dot}~(\ref*{plt:iaj_ji:ai_dot}),
        \algname{ia}{dot}~(\ref*{plt:iaj_ji:ia_dot});
    \item 2 \axpy-based:
        \algname{ij}{axpy}~(\ref*{plt:iaj_ji:ij_axpy}),
        \algname{ji}{axpy}~(\ref*{plt:iaj_ji:ji_axpy});
    \item 2 \gemv-based (see \autoref{alg:iaj_ji}):
        \algname{j}{gemv}~(\ref*{plt:iaj_ji:j_gemv}),
        \algname{i'}{gemv}~(\ref*{plt:iaj_ji:i'_gemv}).
\end{itemize}
The measured and predicted performance for these algorithms is shown in
\autoref{fig:iaj_ji}.  Our predictions clearly discriminate the fastest
algorithm \algname{j}{gemv}~(\ref*{plt:iaj_ji:j_gemv}) across the board.
Furthermore, the next group of four algorithms is also correctly identified and
the low performance (due to the overhead of the involved matrix-copy operation)
of the second \gemv-based algorithm
\algname{i'}{gemv}~(\ref*{plt:iaj_ji:i'_gemv}) is predicted too.

\begin{algorithms}
    \centering\scriptsize
    \newcommand\contraction{iaj_ji}

    \begin{algorithm}{j}{gemv}
        \begin{lstlisting}
for j = 1:$j$
    $C$[:] += $A$[:,:,j]$^T$ $B$[j,:]$^T$!\hfill!(gemv)
        \end{lstlisting}
    \end{algorithm}
    \begin{tikzpicture}[scale=.8]
        \draw[thick, plot1, line cap=rounded, scale=.5] (0,-1,0) -- (0,1,0);
        \node at (0.5,0,0) {$\pluseq$};
        \begin{drawcube}[(1.5,0,0)]
            \sliceC
        \end{drawcube}
        \begin{drawsquare}[(3,0,0)]
            \sliceA
        \end{drawsquare}
    \end{tikzpicture}

    \begin{algorithm}{i'}{gemv}
        \begin{lstlisting}
for i = 1:$i$
    $\widetilde A$[:,:] := $A$[i,:,:]!\hfill!(copy)
    $C$[:]$^T$ += $\widetilde A$[:,:] $B$[:,i]!\hfill!(gemv)
        \end{lstlisting}
    \end{algorithm}
    \begin{tikzpicture}[scale=.8]
        \draw[thick, plot1, line cap=rounded, scale=.5] (0,-1,0) -- (0,1,0);
        \node at (0.5,0,0) {$\pluseq$};
        \begin{drawcube}[(1.5,0,0)]
            \sliceA
        \end{drawcube}
        \begin{drawsquare}[(3,0,0)]
            \sliceB
        \end{drawsquare}
    \end{tikzpicture}

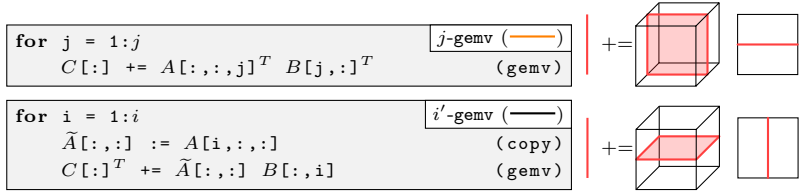
    \caption{Algorithms for $C_a = A_{iaj} B_{ji}$.}
    \label{alg:iaj_ji}
\end{algorithms}

\subsection{Test 3: $C_{abc} = A_{ija} B_{jbic}$, Challenging Contraction}
\label{sec:ijb_jcid}
We now turn to a more complex example: $C_{abc} = A_{ija} B_{jbic}$.  For this
contraction, we look at a total of 176 different algorithms:
\begin{itemize}
    \item 48 \dot-based,
    \item 72 \axpy-based,
    \item 36 \gemv-based,
    \item 12 \ger-based, and
    \item 8 \gemm-based:\\
        \algname{cj'}{gemm}~(\ref*{plt:ijb_jcid:cj'_gemm}),
        \algname{jc'}{gemm}~(\ref*{plt:ijb_jcid:jc'_gemm}),
        \algname{ci'}{gemm}~(\ref*{plt:ijb_jcid:ci'_gemm}),
        \algname{i'c}{gemm}~(\ref*{plt:ijb_jcid:i'c_gemm}),\\
        \algname{bj'}{gemm}~(\ref*{plt:ijb_jcid:bj'_gemm}),
        \algname{jb'}{gemm}~(\ref*{plt:ijb_jcid:jb'_gemm}),
        \algname{bi'}{gemm}~(\ref*{plt:ijb_jcid:bi'_gemm}),
        \algname{i'b}{gemm}~(\ref*{plt:ijb_jcid:i'b_gemm}).
\end{itemize}
All \gemm-based (see \autoref{algs:ijb_jcid}) and several of the \gemv-based
algorithms involve copy operations to ensure that each matrix has a contiguously
stored dimension, as required by the BLAS interface.  Once again, we consider a
very challenging scenario where both contracted indices are of size $i = j = 8$
and the free indices $a = b = c$ vary together. 

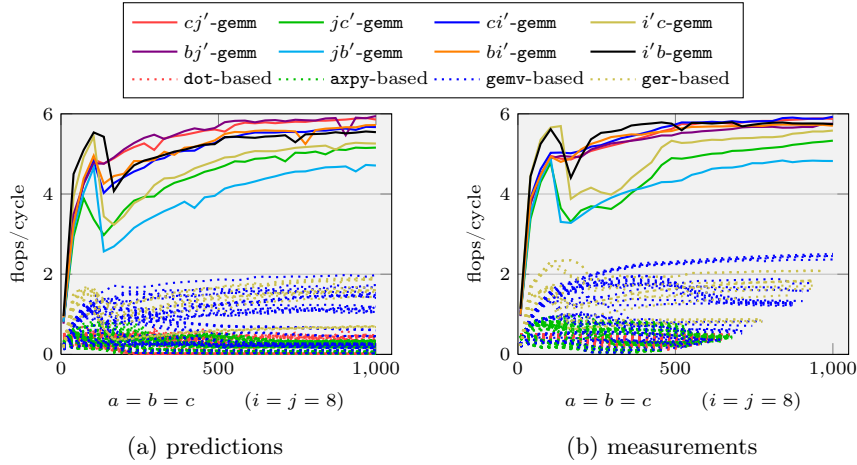
\begin{figure}[t]
    \centering \scriptsize

    \ref*{leg:ijb_jcid}

    \tikzset{external/export=true}

    \begin{subfigure}{.49\textwidth}
        \centering
        \tikzsetnextfilename{ijb_jcid_pred}
        \begin{tikzpicture}
            \begin{axis}[perfplot,
                    twocolplot,
                    cycle list name=default,
                    ymax=6,
                    xlabel={$a = b = c$ \quad\quad ($i = j = 8$)},
                    legend to name=leg:ijb_jcid,
                    legend columns=4,
                ]
                \foreach \var/\loops/\kernel in {%
                    169/cj'/gemm,%
                    170/jc'/gemm,%
                    171/ci'/gemm,%
                    172/i'c/gemm,%
                    173/bj'/gemm,%
                    174/jb'/gemm,%
                    175/bi'/gemm,%
                    176/i'b/gemm%
                    } {
                    \plot file {figures/data/ijb_jcid/pred/var\var.min};
                    \addlegendentryexpanded{\algname{\loops}{\kernel}}
                    \label{plt:ijb_jcid:\loops_\kernel}
                }
                \addlegendimage{plotdot, dotted}
                \addlegendentry{{\tt dot}-based}
                \addlegendimage{plotaxpy, dotted}
                \addlegendentry{{\tt axpy}-based}
                \addlegendimage{plotgemv, dotted}
                \addlegendentry{{\tt gemv}-based}
                \addlegendimage{plotger, dotted}
                \addlegendentry{{\tt ger}-based}
                \foreach \var in {1,...,48}
                    \addplot[plotdot, dotted] file {figures/data/ijb_jcid/pred/var\var.min};
                \foreach \var in {49,...,120}
                    \addplot[plotaxpy, dotted] file {figures/data/ijb_jcid/pred/var\var.min};
                \foreach \var in {121,...,156}
                    \addplot[plotgemv, dotted] file {figures/data/ijb_jcid/pred/var\var.min};
                \foreach \var in {157,...,168}
                    \addplot[plotger, dotted] file {figures/data/ijb_jcid/pred/var\var.min};
            \end{axis}
        \end{tikzpicture}
        \caption{predictions}
        \label{fig:ijb_jcid_pred}
    \end{subfigure}
    \begin{subfigure}{.49\textwidth}
        \centering
        \tikzsetnextfilename{ijb_jcid_meas}
        \begin{tikzpicture}
            \begin{axis}[perfplot,
                    twocolplot,
                    cycle list name=default,
                    ymax=6,
                    xlabel={$a = b = c$ \quad\quad ($i = j = 8$)},
                ]
                \foreach \var in {169,...,176}
                    \plot file {figures/data/ijb_jcid/meas/var\var.min};
                \foreach \var in {1,...,48}
                    \addplot[plotdot, dotted] file {figures/data/ijb_jcid/meas/var\var.min};
                \foreach \var in {49,...,120}
                    \addplot[plotaxpy, dotted] file {figures/data/ijb_jcid/meas/var\var.min};
                \foreach \var in {121,...,156}
                    \addplot[plotgemv, dotted] file {figures/data/ijb_jcid/meas/var\var.min};
                \foreach \var in {157,...,168}
                    \addplot[plotger, dotted] file {figures/data/ijb_jcid/meas/var\var.min};
                \end{axis}
        \end{tikzpicture}
        \caption{measurements}
        \label{fig:ijb_jcid_meas}
    \end{subfigure}

    \tikzset{external/export=false}
    \caption{
        $C_{abc} \coloneqq A_{ija} B_{jbic}$: Performance prediction and
        measurements.
    }
    \label{fig:ijb_jcid}
\end{figure}

Starting with the predictions, in \autoref{fig:ijb_jcid_pred}, we present the
expected flops/cycle of the 176 algorithms, where BLAS-1 and BLAS-2 algorithms
are grouped by kernel.  Even with the copy operations, the \gemm-based
algorithms are the fastest ones.  However, within these 8 algorithms, the
performance differs by more than 20\%.  Focusing on the \gemm-algorithms, we
compare with corresponding performance measurements\footnote{%
    Slow tensor contraction algorithms were stopped before reaching the largest
    test-cases by limiting the total measurement time per algorithm to 15
    minutes.
} in \autoref{fig:ijb_jcid_meas}.  The comparison shows that our predictions
clearly separate the bulk of fast algorithms from the slightly less efficient
ones.

\begin{algorithms}
    \centering\scriptsize
    \newcommand\contraction{ijb_jcid}

    \begin{algorithm}{cj'}{gemm}
         \begin{lstlisting}
for c = 1:$c$
    for j = 1:$j$
        $\widetilde B$[:,:] := $B$[j,:,:,c]!\hfill!(copy)
        $C$[:,:,c] += $A$[:,j,:]$^T$ $\widetilde B$[:,:]$^T$!\hfill!(gemm)
         \end{lstlisting}
    \end{algorithm}
 
    \begin{algorithm}{jc'}{gemm}
         \begin{lstlisting}
for j = 1:$j$
    for c = 1:$c$
        $\widetilde B$[:,:] := $B$[j,:,:,c]!\hfill!(copy)
        $C$[:,:,c] += $A$[:,j,:]$^T$ $\widetilde B$[:,:]$^T$!\hfill!(gemm)
         \end{lstlisting}
    \end{algorithm}
 
    \begin{algorithm}{ci'}{gemm}
         \begin{lstlisting}
for c = 1:$c$
    for i = 1:$i$
        $\widetilde A$[:,:] := $A$[i,:,:]!\hfill!(copy)
        $C$[:,:,c] += $\widetilde A$[:,:]$^T$ $B$[:,:,i,c]!\hfill!(gemm)
         \end{lstlisting}
    \end{algorithm}
 
    \begin{algorithm}{i'c}{gemm}
         \begin{lstlisting}
for i = 1:$i$
    $\widetilde A$[:,:] := $A$[i,:,:]!\hfill!(copy)
    for c = 1:$c$
        $C$[:,:,c] += $\widetilde A$[:,:]$^T$ $B$[:,:,i,c]!\hfill!(gemm)
         \end{lstlisting}
    \end{algorithm}
 
    \begin{algorithm}{bj'}{gemm}
         \begin{lstlisting}
for b = 1:$b$
    for j = 1:$j$
        $\widetilde B$[:,:] := $B$[j,b,:,:]!\hfill!(copy)
        $C$[:,b,:] += $A$[:,j,:]$^T$ $\widetilde B$[:,:]!\hfill!(gemm)
         \end{lstlisting}
    \end{algorithm}
 
    \begin{algorithm}{jb'}{gemm}
         \begin{lstlisting}
for j = 1:$j$
    for b = 1:$b$
        $\widetilde B$[:,:] := $B$[j,b,:,:]!\hfill!(copy)
        $C$[:,b,:] += $A$[:,j,:]$^T$ $\widetilde B$[:,:]!\hfill!(gemm)
         \end{lstlisting}
    \end{algorithm}
 
    \begin{algorithm}{bi'}{gemm}
         \begin{lstlisting}
for b = 1:$b$
    for i = 1:$i$
        $\widetilde A$[:,:] := $A$[i,:,:]!\hfill!(copy)
        $C$[:,b,:] += $\widetilde A$[:,:]$^T$ $B$[:,b,i,:]!\hfill!(gemm)
         \end{lstlisting}
    \end{algorithm}

    \begin{algorithm}{i'b}{gemm}
         \begin{lstlisting}
for i = 1:$i$
    $\widetilde A$[:,:] := $A$[i,:,:]!\hfill!(copy)
    for b = 1:$b$
        $C$[:,b,:] += $\widetilde A$[:,:]$^T$ $B$[:,b,i,:]!\hfill!(gemm)
         \end{lstlisting}
    \end{algorithm}
    
    \caption{
        $C_{abc} = A_{ija} B_{jbic}$, \gemm-based.
    }
    \label{algs:ijb_jcid}
\end{algorithms}

\subsubsection{Multithreading.}
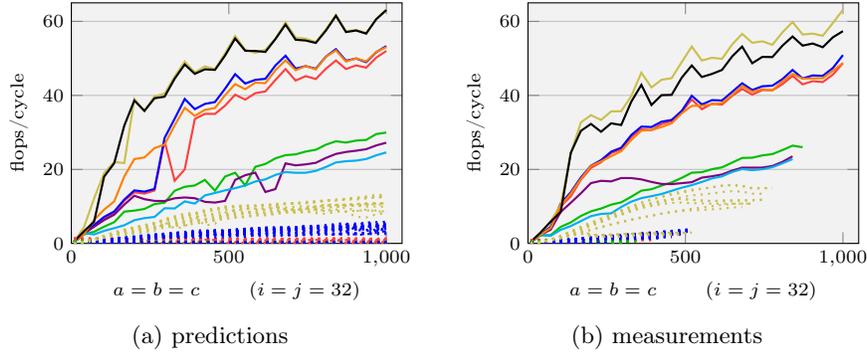
\begin{figure}[t]
    \centering \scriptsize



    \tikzset{external/export=true}

    \begin{subfigure}{.49\textwidth}
        \centering
        \tikzsetnextfilename{ijb_jcid10_pred}
        \begin{tikzpicture}
            \begin{axis}[perfplot,
                    twocolplot,
                    cycle list name=default,
                    ymax=65,
                    xlabel={$a = b = c$ \quad\quad ($i = j = 32$)},
                ]
                \foreach \var in {169,...,176}
                    \plot file {figures/data/ijb_jcid10/pred/var\var.min};
                \foreach \var in {1,...,48}
                    \addplot[plotdot, dotted] file {figures/data/ijb_jcid10/pred/var\var.min};
                \foreach \var in {49,...,120}
                    \addplot[plotaxpy, dotted] file {figures/data/ijb_jcid10/pred/var\var.min};
                \foreach \var in {121,...,156}
                    \addplot[plotgemv, dotted] file {figures/data/ijb_jcid10/pred/var\var.min};
                \foreach \var in {157,...,168}
                    \addplot[plotger, dotted] file {figures/data/ijb_jcid10/pred/var\var.min};
            \end{axis}
        \end{tikzpicture}
        \caption{predictions}
        \label{fig:ijb_jcid10_pred}
    \end{subfigure}
    \begin{subfigure}{.49\textwidth}
        \centering
        \tikzsetnextfilename{ijb_jcid10_meas}
        \begin{tikzpicture}
            \begin{axis}[perfplot,
                    twocolplot,
                    cycle list name=default,
                    ymax=65,
                    xlabel={$a = b = c$ \quad\quad ($i = j = 32$)},
                ]
                \foreach \var in {169,...,176}
                    \plot file {figures/data/ijb_jcid10/meas/var\var.min};
                \foreach \var in {1,...,48}
                    \addplot[plotdot, dotted] file {figures/data/ijb_jcid10/meas/var\var.min};
                \foreach \var in {49,...,120}
                    \addplot[plotaxpy, dotted] file {figures/data/ijb_jcid10/meas/var\var.min};
                \foreach \var in {121,...,156}
                    \addplot[plotgemv, dotted] file {figures/data/ijb_jcid10/meas/var\var.min};
                \foreach \var in {157,...,168}
                    \addplot[plotger, dotted] file {figures/data/ijb_jcid10/meas/var\var.min};
                \end{axis}
        \end{tikzpicture}
        \caption{measurements}
        \label{fig:ijb_jcid10_meas}
    \end{subfigure}

    \tikzset{external/export=false}
    \caption{
        $C_{abc} \coloneqq A_{ija} B_{jbic}$: Performance prediction and
        measurements with 10 threads.
    }
    \label{fig:ijb_jcid10}
\end{figure}

The algorithms considered here can make use of shared memory parallelism by
employing multithreaded BLAS kernels.  To focus on the impact of parallelism, we
increase the contracted tensor dimension sizes to $i = j = 32$ and use all 10
cores of the Ivy Bridge-EP CPU with {\sc OpenBLAS}.  Performance predictions and
measurements for this setup are presented in \autoref{fig:ijb_jcid10}.  Our
predictions correctly separate the three groups of \gemm-based implementations;
moreover, algorithms \algname{i'c}{gemm}~(\ref*{plt:ijb_jcid:i'c_gemm}) and
\algname{i'b}{gemm}~(\ref*{plt:ijb_jcid:i'b_gemm}) (see
\autoref{algs:ijb_jcid}), which reach 60 flops/cycle,\footnote{%
    Using 10 cores, the theoretical peak performance is 80 flops/cycle.
} are identified as the fastest.  The slowest algorithm
(\algname{jb'}{gemm}~(\ref*{plt:ijb_jcid:jb'_gemm})) on the other hand merely
reaches 20 flops/per cycle.  This $3\times$~difference in performance among
\gemm-based algorithms emphasizes the importance of selecting the right
algorithm.

\subsection{Efficiency Study}
\begin{figure}[t]
    \centering \scriptsize

    \tikzset{external/export=true}
    \tikzsetnextfilename{eff}
    \begin{tikzpicture}
        \begin{semilogyaxis}[
                ymin=1e1,
                ymax=1e9,
                xlabel={$a = b = c$ \quad $(i = 8)$},
                ylabel={time(execution) / time(benchmark)},
                legend columns=-1,
                legend pos=south east,
            ]
            \addlegendimage{empty legend}
            \addlegendentry{kernel:}
            \addlegendimage{plot1}
            \addlegendentry{{\tt dot}}
            \label{plt:eff:dot}
            \addlegendimage{plot2}
            \addlegendentry{{\tt axpy}}
            \label{plt:eff:axpy}
            \addlegendimage{plot3}
            \addlegendentry{{\tt gemv}}
            \label{plt:eff:gemv}
            \addlegendimage{plot4}
            \addlegendentry{{\tt ger}}
            \label{plt:eff:ger}
            \addlegendimage{plot5}
            \addlegendentry{{\tt gemm}}
            \label{plt:eff:gemm}
            \foreach \var in {1,...,6}
                \plot[plot1] file {figures/data/eff/var\var.dat};
            \foreach \var in {7,...,24}
                \plot[plot2] file {figures/data/eff/var\var.dat};
            \foreach \var in {25,...,30}
                \plot[plot3] file {figures/data/eff/var\var.dat};
            \foreach \var in {31,...,34}
                \plot[plot4] file {figures/data/eff/var\var.dat};
            \foreach \var in {35,...,36}
                \plot[plot5] file {figures/data/eff/var\var.dat};
        \end{semilogyaxis}
    \end{tikzpicture}

    \tikzset{external/export=false}
    \caption[]{
        $C_{abc} \coloneqq A_{ai} B_{ibc}$: Prediction efficiency.
    }
    \label{fig:eff}
\end{figure}
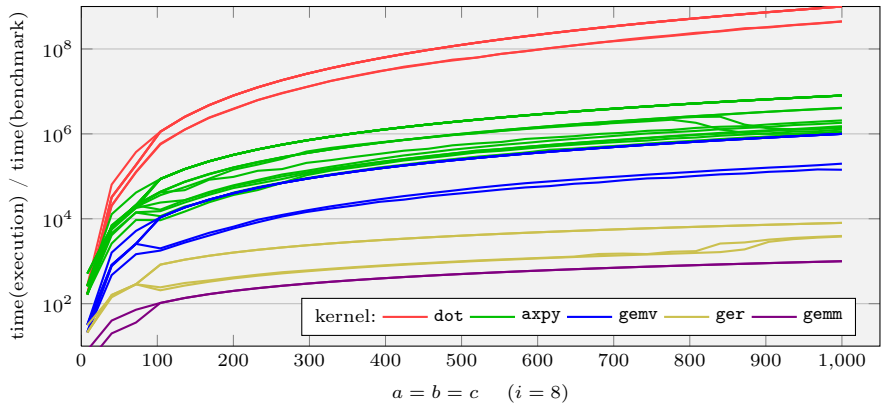

\parsum{compare total prediction time with measurement time}
The ultimate goal of this work is to automatically and quickly select the
fastest algorithm for a given tensor contraction. The experiments presented so
far provide evidence that our automated approach successfully identifies the
fastest algorithm(s).  With this last experiment, we investigate the efficiency
of our micro-benchmark-based approach.  For this purpose, we use again the
contraction $C_{abc} = A_{ai} B_{ibc}$, with $i = 8$ and varying $a = b = c$.
\hyperref[fig:eff]{Figure~\ref*{fig:eff}} displays the ratio of how much faster
our micro-benchmark is compared to executing the corresponding algorithm.  In
general, our prediction proves to be several orders of magnitude faster than the
algorithm itself.  At $a = b = c = 1{,}000$, this relative improvement is
smallest for the \gemm-based algorithms~(\ref*{plt:eff:gemm}) at $10^3\times$,
since each \gemm performs a significant portion of the computation; for the
\ger-based algorithms~(\ref*{plt:eff:ger}), it lies between $6\cdot10^3$ and
$10^4\times$ and for the \gemv-based algorithms~(\ref*{plt:eff:gemv}) the gain
is $5\cdot10^5$ to $10^6\times$; finally, the gain for both BLAS-1-based
algorithms~(\ref*{plt:eff:axpy}, \ref*{plt:eff:dot}), where each BLAS-call only
performs a tiny fraction of the contraction, our prediction is between $6$ and
$9$ orders of magnitude faster than the execution.

    \section{Conclusion}
    In this paper, we focused on the performance prediction of BLAS-based algorithms
for tensors contractions.  First, based on previous work, we developed an
algorithm and code generator that given the mathematical description of a tensor
contraction, casts the computation in terms of one of five different BLAS
kernels; since, in general, a tensor contraction may be decomposed in terms of
matrix and vector products in many different ways, the generator often returns
dozens of alternative algorithms.

Then, we tackled the problem of selecting the fastest algorithms without ever
executing them. Instead of executing the full algorithms, our approach is based
on timing the BLAS kernels in a small set of micro-benchmarks.  These
micro-benchmarks are run in a context that emulates that of the actual
computation; thanks to a careful treatment of cache-locality and a model of the
cache prefetcher's behavior, our performance prediction tool is capable of
identifying the best-performing algorithms in a tiny fraction of the time
required to actually run and time all of them.

The quality of the predictions was showcased for a number of challenging
scenarios, including contractions among tensors with small dimensions,
contractions that can only be cast in terms of BLAS 1 and BLAS 2 kernels, and
multi-threaded computations.


    \bibliographystyle{splncs}
    \bibliography{references}

\begin{thebibliography}{10}

\bibitem{blas1}
Lawson, C.L., Hanson, R.J., Kincaid, D.R., Krogh, F.T.:
\newblock Basic linear algebra subprograms for fortran usage.
\newblock ACM Trans. Math. Softw. \textbf{5}(3) (September 1979)  308--323

\bibitem{blas2}
Dongarra, J.J., Du~Croz, J., Hammarling, S., Hanson, R.J.:
\newblock An extended set of fortran basic linear algebra subprograms.
\newblock ACM Trans. Math. Softw. \textbf{14}(1) (March 1988)  1--17

\bibitem{blas3}
Dongarra, J.J., Du~Croz, J., Hammarling, S., Duff, I.S.:
\newblock A set of level 3 basic linear algebra subprograms.
\newblock ACM Trans. Math. Softw. \textbf{16}(1) (March 1990)  1--17

\bibitem{tce}
Baumgartner, G., Auer, A., Bernholdt, D., Bibireata, A., Choppella, V.,
  Cociorva, D., Gao, X., Harrison, R., Hirata, S., Krishnamoorthy, S.,
  Krishnan, S., Lam, C., Lu, Q., Nooijen, M., Pitzer, R., Ramanujam, J.,
  Sadayappan, P., Sibiryakov, A.:
\newblock Synthesis of high-performance parallel programs for a class of ab
  initio quantum chemistry models.
\newblock Proceedings of the IEEE \textbf{93}(2) (Feb 2005)  276--292

\bibitem{lu}
Lu, Q., Gao, X., Krishnamoorthy, S., Baumgartner, G., Ramanujam, J.,
  Sadayappan, P.:
\newblock Empirical performance model-driven data layout optimization and
  library call selection for tensor contraction expressions.
\newblock J. Parallel Distrib. Comput. \textbf{72}(3) (March 2012)  338--352

\bibitem{DiNapoli2014:210}
{Di Napoli}, E., Fabregat-Traver, D., Quintana-Orti, G., Bientinesi, P.:
\newblock Towards an efficient use of the blas library for multilinear tensor
  contractions.
\newblock Applied Mathematics and Computation \textbf{235} (May 2014)  454--468

\bibitem{roman}
Iakymchuk, R., Bientinesi, P.:
\newblock Modeling performance through memory-stalls.
\newblock SIGMETRICS Perform. Eval. Rev. \textbf{40}(2) (October 2012)  86--91

\bibitem{roman2}
Iakymchuk, R., Bientinesi, P.:
\newblock Execution-less performance modeling.
\newblock In: Proceedings of the Second International Workshop on Performance
  Modeling, Benchmarking and Simulation of High-Performance Computing Systems
  (PMBS11) held as part of the Supercomputing Conference (SC11), Seattle, USA
  (November 2011)

\bibitem{modeling}
Peise, E., Bientinesi, P.:
\newblock Performance modeling for dense linear algebra.
\newblock In: Proceedings of the 2012 SC Companion: High Performance Computing,
  Networking Storage and Analysis. SCC '12, Washington, DC, USA, IEEE Computer
  Society (2012)  406--416

\bibitem{openblas}
OpenBLAS:
\newblock \url{http://xianyi.github.com/OpenBLAS}

\end{thebibliography}
\end{document}